# Dust traps and gas kinematic signatures in a crescent structure of a planet-forming disk

Greta Guidi[1]*, François Menard[1], Daniel J. Price[1,2], Marion Villenave[1] and Jie Ma[1]

[1]University Grenoble Alpes, CNRS, IPAG, Grenoble, France, [2]School of Physics and Astronomy, Monash University, Melbourne, VIC, Australia

In the past few years, ALMA unveiled a variety of substructures (rings, spirals, crescents) in the continuum emission of most protoplanetary disks imaged at high spatial resolution. While the majority of disks presents axisymmetric ring-like structures in the dust brightness distribution, some sources display asymmetric morphologies (blobs, crescents) that have been often associated to vortices and/or mechanisms generated by the presence of one or more embedded planets. In this brief research report we present the analysis of the arc structure observed in the dust continuum emission of the disk around HD 163296, using high resolution (~8 au) matched continuum data from ALMA at four wavelengths. We characterize in detail the arc structures and present a kinematic signature observed in the CS(3−2) emission at the same location. Our results indicate that the crescent is caused by differential dust trapping in a local pressure maxima, for which plausible mechanisms can be the presence of a vortex or trapping in a Lagrangian point of the planet-star system.

KEYWORDS
crescent, CS, HD 163296, planet-disk interaction, protoplanetary disk, vortex

## 1 Introduction

Protoplanetary disks of gas and dust are naturally formed during the star formation process from the collapse of molecular clouds. With a typical lifetime of 1–10 Myr (e.g., Mamajek, 2009) they are regarded as the primary environment for planet formation. The advent of a new generation of observatories and especially ALMA (Atacama Large Millimeter/submillimeter Array) has enabled the observation of disks with exquisite detail, unveiling small-scale structures in the dust and gas distributions, such as concentric rings, spirals, lopsided patterns. Multiple mechanisms have been invoked to explain such features, such as vortices forming after the development of dynamical instabilities (e.g., Flock et al., 2015; Surville and Barge, 2015), presence of embedded planets (e.g., Dipierro et al., 2015), or chemical effects at the locations of snowlines (e.g., Zhang et al., 2015).

In this short research report, we present the analysis of the crescent structure observed in the disk around the Herbig Ae star HD 163296 (MWC 275) at four different ALMA bands. A non-symmetrical crescent structure in the 1.3 mm HD 163296 continuum emission was characterised in Isella et al. (2018). After subtracting an axysymmetric model, Isella et al. (2018) found a radius of 55 au and a position angle of 141.8° (East from North) for the crescent. Here we want to investigate the nature of this structure by studying its appearance at several wavelengths. We characterize its azimuthal morphology and show the presence of a kinematic signature in the gas at the same location. We compare our findings to theoretical predictions of disk instabilities and planet-disk interaction in order to give an interpretation of the mechanisms producing such asymmetric structure.

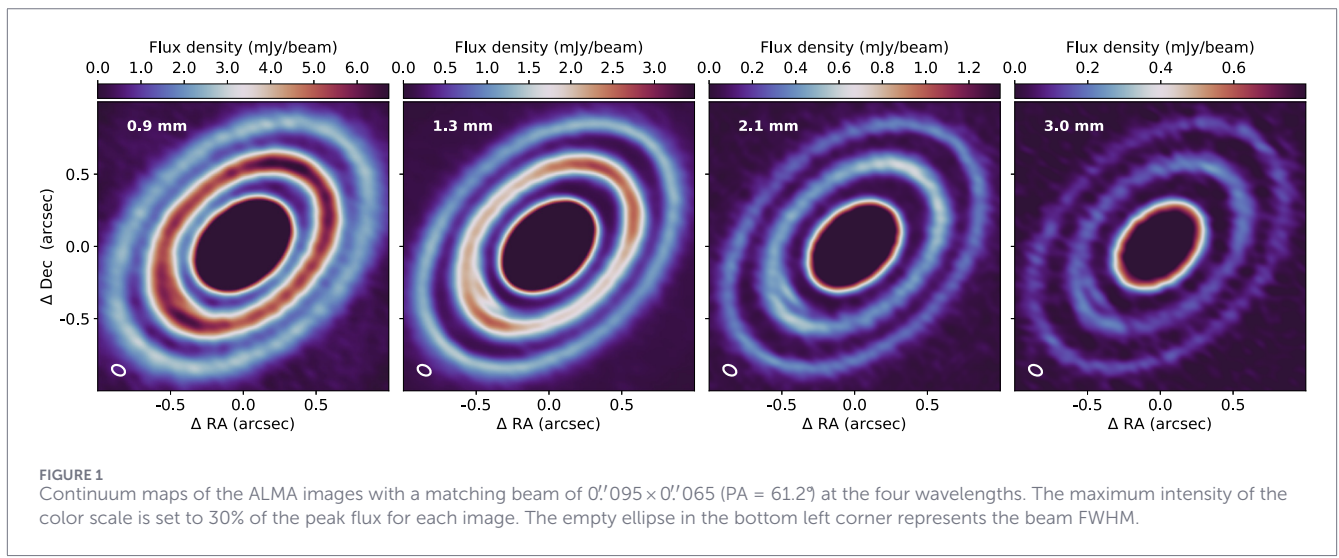

**FIGURE 1**
Continuum maps of the ALMA images with a matching beam of 0.″095 × 0.″065 (PA = 61.2°) at the four wavelengths. The maximum intensity of the color scale is set to 30% of the peak flux for each image. The empty ellipse in the bottom left corner represents the beam FWHM.

## 2 Methods

### 2.1 Image analysis

We analysed high resolution ALMA continuum observations of HD 163296 at four different bands (Band 3, 4, 6 and 7), displayed in Figure 1, and the CS(3–2) emission line detected at Band 4. A description of the observations and the imaging procedure is given in the Supplementary Material. We perform a fit in the image plane to derive the geometrical parameters such as inclination, position angle and radial separation of the crescent and its neighboring ring, we refer to the Supplementary Material for a more detailed description of the fitting procedure. We find that the radial separation for the crescent is consistent across the four wavelength, with an average value of 0.″56 (or 56.7 au at a distance of 101.2 pc, Bailer-Jones et al., 2018), while the azimuthal extent varies with wavelength (see Section 3).

### 2.2 SED fitting

We model the azimuthal profiles of the crescent with a simple radiative transfer analytical description, where we describe the optical depth as a power law as function of frequency. Under the assumption of LTE (Local Thermodynamic Equilibrium), the dust thermal emission can be described as in Equation 1:

$$I_\nu(r) = B_\nu(T(r)) \cdot \left(1 - e^{-\tau_\nu(r)/\mu}\right), \quad (1)$$

where $\tau_\nu = \tau_0 (\nu/\nu_0)^\beta$ is the optical depth, $\mu = \cos(i)$ is the inclination parameter and $B_\nu(T(r))$ is the Planck function. We use Monte Carlo nested sampling algorithm in UltraNest (Buchner, 2021) to fit the optical depth $\tau$, and the spectral index $\beta$ across the crescent structure. We use a value of 20 K for the Temperature at the crescent position, taken from the SED modeling of the radial intensity profiles performed in Guidi et al. (2022). In addition, we perform the same fit using T = 15 K and T = 40 K. The error associated to each data point includes both the statistical error (rms of the clean image) and the systematic error associated to flux calibration.

## 3 Results

We show in Figure 2 the azimuthal profiles of the crescent structure and the results of the gaussian fit at each wavelength. The peak is found to be around 120° on average, but we note a small progressive increase of the peak position angle with wavelength, with about 3° separation between the 0.9 mm peak and the 3 mm peak. Another clear feature is the decrease of the gaussian width with wavelength, with a FWHM (Full Width Half Maximum) of about 42 au at 0.9 mm to 32 au at 3 mm. Finally, we note the presence of a double peak only in the 3 mm profiles. The distance between the two peaks is about 15° (or 14 au), although this value is affected by our angular resolution and the location of the azimuthal bins. We fit the peak position and the crescent FWHM from the gaussian fitting with a power law as a function of wavelength $a \cdot \lambda^p$, and we obtain coefficients of p = −0.01±0.002 and p = −0.21±0.02, respectively (see Figure 2, bottom panel). We use the python `scipy.curve_fit` function and estimated the errors on the coefficient from the square-root of the covariance matrix.

The result of the parametric SED fitting is shown in Figure 3: the emission results moderately optically thin across the analysed spectrum, with a peak value of 0.74 at 0.9 mm, while at 3 mm $\tau$ reaches a value of about 0.14. For comparison, we perform the same fit using T = 15 K and T = 40 K in order to have a reasonable interval in dust temperature: we show the results with dashed line. If the dust temperature was higher than the fixed value of 20 K, the resulting optical depth would be smaller, while a lower T of 15 K gives a optical depth slightly lower than 1 at the crescent peak. The dust opacity spectral index $\beta$ is consistent with a value of 1.7 (typically measured in the ISM and indicating sub-micron sized grains) at most azimuthal direction, decreasing only between 100° and 150° (i.e., at the location of the crescent), and in this range it reaches an average value of 1.3.

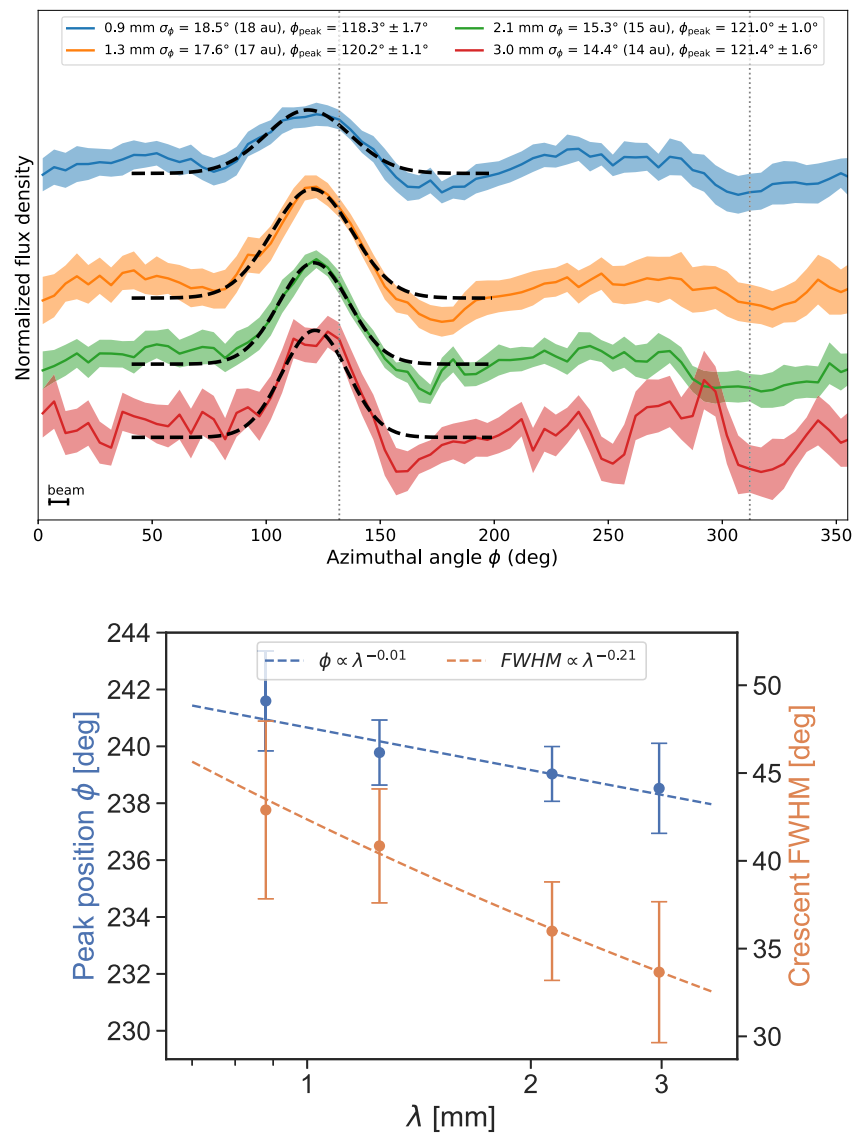

FIGURE 2
Top: Deprojected azimuthal profiles (PA = 132°, incl = 45.3°) at a separation of 0.″56 for the 4 ALMA wavelengths, each normalized to their peak value and offset by 1.2 for better visibility. The azimuthal angle starts from North and goes counter-clockwise. The bins are computed every 5° (corresponding to half a beam) and the error bars are the sum in quadrature of the standard deviation of the points within each bin and the maps rms errors. Vertical dotted lines are drawn at the position of the disk major axis. Bottom: Peak position and crescent width as a function of wavelength, with power-law fits in dashed lines. The peak position is described as 360-$\phi_{peak}$, so that the azimuth increases in the clockwise direction, consistent with the disk rotation.

An inspection of the velocity (moment 1) and velocity dispersion (moment 2) maps of the CS (3–2) line reveals a distortion in velocity at the location of the dust crescent. In particular, a local increase in velocity of ~300 m/s with respect to the neighbouring regions is observed in the moment 1 map, highlighting motion towards the midplane. At the same location, we see a local enhancement of the velocity dispersion, with a maximum value of 3.5 km/s (see Figure 4). The projected separation of the peak in velocity dispersion is 0.″52 at a position angle of 135°. The peak flux in the CS cube at this position is 5.0±0.2 mJy/beam, and the line to continuum ratio is about 6, where to compute the ratio we smoothed the continuum map at 2 mm to the beam of the CS cube (0.″165 × 0.″120, PA −65.17°).

# 4 Discussion

Dust asymmetric small-scale structures have been observed with ALMA at high spatial resolution (≤10 au) and at multiple wavelengths only in a handful of disks. Two notable examples are HD 135344B (SAO 206462), observed at band 9, 7, 4 and 3 by Cazzoletti et al. (2018), and HD 163296, presented in this work. The dust emission morphology in these two disks present some common features, namely, a shift in the azimuthal location of the peak flux for different wavelengths, a decreasing azimuthal extent with increasing wavelength, and a multi-peaked emission at the longer wavelengths (typically 3 mm). The almost flat coefficient of −0.01 that we find for the azimuthal shift of the crescent peak as a function of wavelength, reflects the small offset of a few degrees that is observed across our wavelength range. Nevertheless, the shift is monotonic and oriented opposite to the disk rotation. We note that in Cazzoletti et al. (2018) the azimuthal shift of the peak was significant (50° between 0.4 mm and 3 mm), but in both cases the direction of the shift is opposite to what predicted by the effect of self gravity in the presence of vortex, as shown by Baruteau and Zhu (2016): in their simulations grains with Stokes number much smaller and much larger than unity are trapped near the vortex center, while particles with St~1 are shifted ahead of the vortex. The decreasing azimuthal extent of the dust crescent with wavelength is consistent with size-differential dust trapping, with larger grains being more effectively retained at the pressure maxima. This behaviour is predicted by both trapping in a vortex (e.g., Birnstiel et al., 2013; Baruteau and Zhu, 2016) or in a Lagrangian point (Rodenkirch et al., 2021). The optical depth can also partially contribute to this effect, since dust emission becomes more optically thin at longer wavelengths (see previous Section).

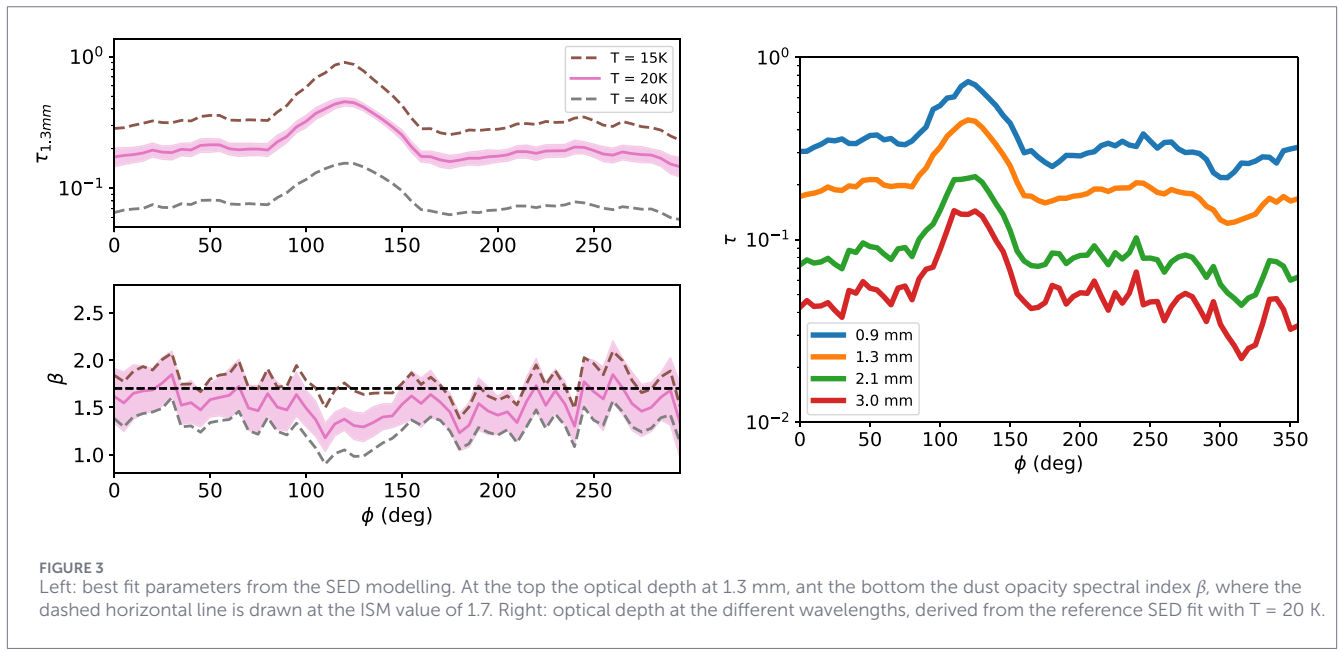

FIGURE 3
Left: best fit parameters from the SED modelling. At the top the optical depth at 1.3 mm, ant the bottom the dust opacity spectral index $\beta$, where the dashed horizontal line is drawn at the ISM value of 1.7. Right: optical depth at the different wavelengths, derived from the reference SED fit with T = 20 K.

Another interesting aspect is that at the same resolution where a single peak in the azimuthal direction is observed at most ALMA bands, 3 mm data show a double or multiple peak emission. Different mechanisms have been proposed to explain this feature: it has been shown that dust-feedback in a vortex can produce multiple azimuthal overdensities in the dust distribution. Vortices formed at the edge of dead zones can produce multiple small clumps of dust that, if not resolved, appear as a single feature (Miranda et al., 2017). This mechanism would also produce a axisymmetric dust ring at a radial distance twice the one of the asymmetric structure. Also, vortices can be formed at the edge of a gap opened by a forming planet, as a result of RWI (Rossby Wave Instability). These are generally short-lived, but can survive for a longer time if both the dust-to-gas ratio and the viscosity are very low (e.g., Fu et al., 2014; Surville et al., 2016). Baruteau and Zhu, (2016) show that grains with different stokes number peak at a shifted azimuthal angle, and a double azimuthal peak in optical depth can be formed for a p = 2.5 dust size distribution.

Other mechanisms do not require the presence of a vortex, and in the case of HD 163296, the presence of one or multiple planets within the dust gaps has been suggested by multiple works (e.g., Isella et al., 2016; Isella et al., 2018), when modelling the dust and molecular gas emission. A few studies focused on reproducing the asymmetric crescent: Rodenkirch et al. (2021) performed a multi-fluid simulation for this disk including three planets at separations of 48, 83, and 137 au, and found that a 1 $M_{Jup}$ planet orbiting in the gap at a radius of 48 au could reproduce the shape (but not the location) of the crescent, that would result from trapping of dust particles at its Lagrangian point L5. Garrido-Deutelmoser et al. (2023) showed instead that a pair of sub-Saturn mass planets in the same gap can reproduce both the shape and the location of the crescent, again corresponding to dust trapping in the L5 Lagrangian point of the outermost planet. An interesting feature seen in these simulations is a small azimuthal shift for particles with different Stokes number (grains with St~0.1 would be closer to the planet compare to grains with ST~$10^{-3}$); while a double peak appears in the dust surface density at L5 when dust feedback is included in the model (see Figures 8, 9 in Rodenkirch et al., 2021). Therefore, trapping in a Lagrangian point could be consistent with the features we identify in the dust distribution as shown in this study. However, vortex-induced trapping of dust could not be ruled out.

Insights on the gas dynamics could help disentangle the origin of the crescent: the CS molecule, with its lower abundance (hence lower optical depth) with respect to CO, is thought to come from closer to the midplane, and it is a sensitive tracer of non-thermal broadening due to its larger molecular weight. The compactness of the CS kinematic deviation and its co-location with the dust trapping crescent would be more consistent with the presence of a vortex, since the kinematic signatures in case of embedded planets would be more extended in area when corresponding to spiral wakes (Pinte et al., 2019) or more offsets towards the position of the planet (e.g., Pinte et al., 2023). We note that the low signal-to-noise makes it difficult to pinpoint the scale height of the CS emission at the different radii and subtract a Keplerian model (see Law et al., 2025), so that the exact location of the velocity kink projected on the midplane is uncertain. Regardless, both the positive direction of the velocity distortion (see Figure 4, top panel), and the higher velocity dispersion, indicate a vertical motion of the gas towards the midplane. Such vertical gas flow is predicted by hydrodynamic simulation of disk-planet interaction, and would be located at the position of a putative protoplanet (Szulágyi et al., 2022). We note that a higher gas velocity dispersion is also predicted in the presence of a vortex generated by RWI (e.g., Huang et al., 2018).

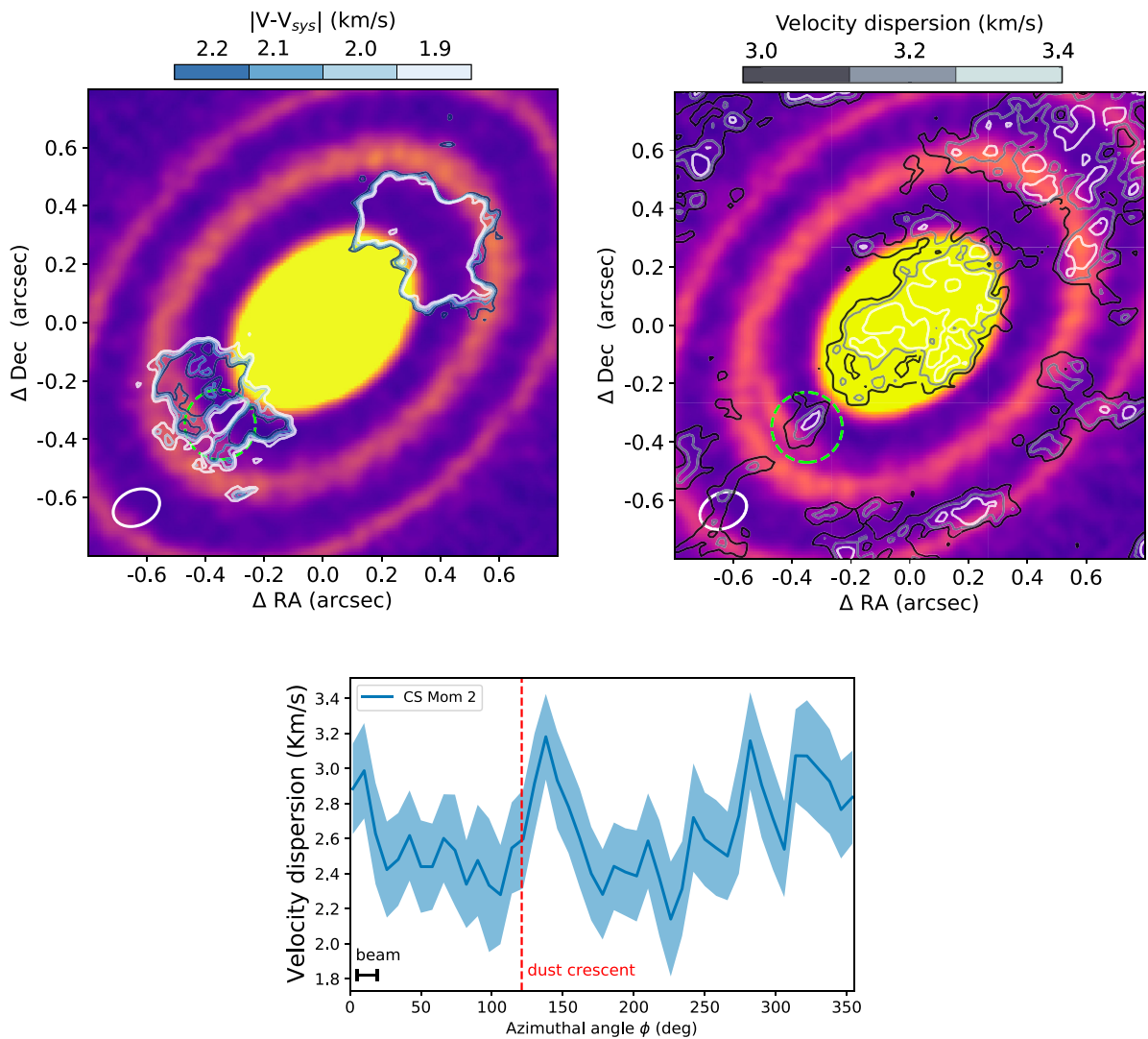

FIGURE 4
Top panels: in color scale, continuum emission at 2 mm. Left: contours of the moment 1 of CS(3–2) drawn at a Δv = ±1.9 to ±2.2 km/s from the system velocity $v_{sys}$ = 5.3 km/s. Right: contours of velocity dispersion (moments 2 of CS (3–2)), drawn every 0.2 km/s from 3.0 (dark blue) to 3.4 Km/s (white). A localized enhancement of the velocity dispersion is observed in correspondence of the dust crescent. In both panels the local distortions are highlighted by a dashed green circle. *Bottom panel*: azimuthal profile of the CS (3–2) Moment 2, computed with the same geometrical parameters and uncertainties as for the dust crescent (see Figure 2). The peak in velocity dispersion appears slightly shifter with respect to the dust crescent (the average peak position across the four wavelengths is displayed as a dashed red line).

In conclusion, while all the features we identified in the dust crescent in HD 163296 are consistent with dust trapping, it is difficult to disentangle between a vortex or trapping at Lagrangian points of the planet-star system. The inclusion of the CS kinematic signature favors the vortex interpretation, or the presence of a planet at one extremity of the dust crescent. Future high resolution and high sensitivity observations of both longer wavelength continuum emission and molecular tracers will be crucial for disentangling between the mechanisms shaping the HD 16326 morphology.

## 5 Softwares

In this paper we made use of the *CASA* software, the python packages *bettermoments*, *scipy*, *UltraNest*.

## Data availability statement

Publicly available datasets were analyzed in this study. This data can be found here: https://almascience.eso.org/aq/ .

## Author contributions

GG: Conceptualization, Investigation, Writing – original draft, Writing – review and editing. FM: Conceptualization, Funding acquisition, Writing – original draft, Writing – review and editing. DP: Conceptualization, Writing – original draft, Writing – review and editing. MV: Conceptualization, Writing – original draft, Writing – review and editing. JM: Conceptualization, Writing – original draft, Writing – review and editing.

## Funding

The author(s) declared that financial support was received for this work and/or its publication. This project has received funding from the European Research Council (ERC) under the European Union's Horizon Europe research and innovation program (grant agreement No. 101053020, project Dust2Planets).

## Conflict of interest

The author(s) declared that this work was conducted in the absence of any commercial or financial relationships that could be construed as a potential conflict of interest.

## Generative AI statement

The author(s) declared that generative AI was not used in the creation of this manuscript.

Any alternative text (alt text) provided alongside figures in this article has been generated by Frontiers with the support of artificial intelligence and reasonable efforts have been made to ensure accuracy, including review by the authors wherever possible. If you identify any issues, please contact us.

## Publisher's note

All claims expressed in this article are solely those of the authors and do not necessarily represent those of their affiliated organizations, or those of the publisher, the editors and the reviewers. Any product that may be evaluated in this article, or claim that may be made by its manufacturer, is not guaranteed or endorsed by the publisher.

## Supplementary material

The Supplementary Material for this article can be found online at: https://www.frontiersin.org/articles/10.3389/fspas.2026.1727532/full#supplementary-material